\pdfoutput=1 
\documentclass[a4paper, 11pt]{article}
\usepackage[dvipsnames]{xcolor}

\usepackage{jheppub}

\makeatletter
\gdef\@fpheader{}
\makeatother

\usepackage[T1]{fontenc}
\usepackage{amsmath,amssymb,amsthm,amsfonts,graphicx,hyperref}
\usepackage{stmaryrd}
\usepackage[utf8]{inputenc}
\usepackage{bm}

\hypersetup{ linktoc=all,
    colorlinks, linkcolor=blue,
    citecolor=magenta, urlcolor=red
}

\newcommand{\inbox}[1]{\tcbset{fonttitle=\scriptsize} \tcboxmath[colback=white,colframe=black!70]{#1}}

\newcommand{\be}{\begin{equation}}
\newcommand{\ee}{\end{equation}}
\newcommand{\bea}{\begin{eqnarray}}
\newcommand{\eea}{\end{eqnarray}}

\makeatletter \@addtoreset{equation}{section}

\usepackage[most]{tcolorbox}
\tcbset{highlight math style={left=02mm,right=02mm,top=02mm,bottom=02mm}}
\usepackage{empheq}

\begin{document}

\title{\LARGE{\centerline{Null Strings Gauged and Reloaded, II:}} 
\centerline{Consistent  Classical Treatment of the Null Strings}}

\author[a]{M.M. Sheikh-Jabbari}
\author[b]{, H. Yavartanoo}
\affiliation{$^a$ School of Physics, Institute for Research in Fundamental
Sciences (IPM), P.O.Box 19395-5531, Tehran, Iran}
\affiliation{$^b$ Beijing Institute of Mathematical Sciences and Applications (BIMSA), Huairou District, Beijing 101408, P. R. China}
\emailAdd{  
jabbari@theory.ipm.ac.ir, yavar@bimsa.cn}

\abstract{We observed in \cite{Sheikh-Jabbari:2026cnj} that the null strings, tensionless strings with Carrollian worldsheets, exhibit an extra gauge symmetry, \textit{Carroll-Weyl} gauge symmetry, which cannot be obtained from ultra-relativistic Carrollian limit of tensile strings. Due to the existence of this symmetry, the BMS$_3$ algebra of constraints, which is obtained as the Carrollian limit of two Virasoro algebras of the standard tensile strings, should  be replaced with an BMS$_3$ algebra extended by a weight one operator. To establish further the existence and necessity of the Carroll-Weyl gauge symmetry, we carefully work through Hamiltonian analyses of constrained/gauged systems. We also discuss the extended BMS$_3$ algebra of constraints.}  
\maketitle
\flushbottom

\section{Introduction}\label{sec:motivation_generalized_action}

Null string theory has been of interest in the last 4 decades when dealing with certain high energy limit of tensile strings or when strings probing horizons (null surfaces in the target space), see \cite{Bagchi:2026wcu, Sheikh-Jabbari:2026cnj} for more discussions and references therein. In \cite{Sheikh-Jabbari:2026cnj} we made an unexpected observation regarding the Isberg-Lindstr\"om-Sundborg-Theodoridis (ILST) null string theory \cite{Isberg:1992ia, Isberg:1993av}: This action exhibits a (gauge) symmetry that is generated by codimension-1 functions on the worldsheet, in contrast to functions over the whole worldsheet, e.g. as in 2D diffeomorphisms. The existence of this symmetry that was overlooked in the last 33 years of the null string analysis, is surprising and asks for a thorough understanding and discussion. In this and a companion short note \cite{Sheikh-Jabbari:2026vqh}, we take on this task. 

The worldsheet theory of null strings is a 2D Carrollian field theory. The well-known null string action can be obtained through a tensionless limit of ordinary tensile strings \cite{Isberg:1992ia, Isberg:1993av}. This action enjoys Weyl + diffeomorphism gauge symmetry and an extensive literature has been devoted to studying it; see e.g. \cite{Sundborg:1994py, Bagchi:2019cay, Bagchi:2020fpr, Bagchi:2020ats, Bagchi:2021ban} and \cite{Bagchi:2026wcu} for a recent review and references therein. As discussed in \cite{Sheikh-Jabbari:2026vqh}, the ILST action may be viewed as the (partially) gauge-fixed form of a more general action with an addition \textit{Carroll-Weyl} gauge symmetry. The residual part of the Carroll-Weyl gauge symmetry, after gauge-fixing yields the codimension-1 (gauge) symmetry of the ILST action, as noted in \cite{Sheikh-Jabbari:2026cnj}.  The Carroll-Weyl gauge symmetry is specific to Carrollian worldsheets and  has no geometric counterpart on the tensile string worldsheet. 

In this note, we provide a more detailed analysis of the statements made in \cite{Sheikh-Jabbari:2026cnj}.  We perform a detailed Hamiltonian analyses of gauged-systems and establish that the three constraints resulting from the diffeomorphisms + Carroll-Weyl gauge symmetries are required form the consistency of the treatment. We study in more detail the physically inequivalent configurations by solving the three constraints resulting from gauge-fixings, as well as fixing three residual, codimension-1 gauge freedoms. Therefore, showing explicitly that for a null strings with $D$ dimensional Minkowski target space, classically, there are only $D-3$ propagating modes.

\section{Null string action, basic analysis}\label{sec:basic_analysis}

The null string is traditionally described by the Isberg-Lindstr\"om-Sundborg-Theodoridis (ILST) action \cite{Isberg:1992ia, Isberg:1993av}, which is obtained via as a tensionless limit of the standard tensile string in which the worldsheet becomes a 2D Carrollian manifold. Explicitly, 
\begin{equation}\label{ILST}
S_{\text{ILST}} = \frac{\kappa}{2} \int d\tau\,d\sigma \; \mathcal{V}^a \mathcal{V}^b \, \partial_a X^\mu \partial_b X_\mu,
\end{equation}
where $\mathcal{V}^a$ is a worldsheet vector density of weight $+1/2$ and $X^\mu(\tau,\sigma)$ are the embedding fields in a $D$-dimensional target space, which we take to be the Minkowski spacetime with  $\eta_{\mu\nu} = \mathrm{diag}(-1, +1, \cdots, +1)$.  The constant $\kappa$ will be set to $1$ in most of our analysis, though it can be restored by dimensional analysis.
Action \eqref{ILST} is invariant under two sets of local symmetries,  worldsheet diffeomorphisms generated by a vector field $\xi^a(\tau,\sigma)$ and the Carroll-Weyl scalings generated by $\chi(\tau, \sigma)$ subject to ${\cal V}^a\partial_a \chi=0$ \cite{Sheikh-Jabbari:2026cnj}. Denoting the two by $\eta=(\xi^a, \chi)$,  
\begin{align}
\delta_\eta X^\mu &= \xi^a \partial_a X^\mu+ \chi X^\mu, \qquad{\cal V}^a\partial_a \chi=0 \label{X-eta-transf}\\
\delta_\eta \mathcal{V}^a &= \xi^b \partial_b \mathcal{V}^a - \mathcal{V}^b \partial_b \xi^a + \frac{1}{2} (\partial_b \xi^b) \mathcal{V}^a- \chi {\cal V}^a \label{Va-eta-transf}
\end{align}
where $(\partial_b \xi^b) \mathcal{V}^a$ term accounts for the density weight $+1/2$ of $\mathcal{V}^a$. 

As stressed in \cite{Sheikh-Jabbari:2026cnj}, to be a symmetry of \eqref{ILST}, the $\chi$ function in the above is only a function of one direction on the worldsheet. However, one can promote it to a fully fledged, standard, gauge symmetry, by introducing a worldsheet gauge field $\mathcal{W}_a$ (a Carroll-Weyl connection) and replacing ordinary derivatives with covariant derivatives $D_a = \partial_a + \mathcal{W}_a$.  The gauged action becomes \cite{Sheikh-Jabbari:2026vqh}:
\begin{equation}
\inbox{S = \frac{\kappa}{2} \int d\tau\,d\sigma \; \mathcal{V}^a \mathcal{V}^b \, D_a X^\mu D_b X_\mu,}
\label{gauged_action}
\end{equation}
which is invariant under local diff + scaling transformations as in \eqref{X-eta-transf} and \eqref{Va-eta-transf} but with generic $\chi=\chi(\tau,\sigma)$ and,
\begin{equation}\label{Wa-eta-transf}
\delta_\eta \mathcal{W}_a = \xi^b \partial_b \mathcal{W}_a + \mathcal{W}_b \partial_a \xi^b - \partial_a\chi \,.
\end{equation}
A preliminary analysis of this gauged null string action was provided in \cite{Sheikh-Jabbari:2026vqh} where it was argued that the classical dynamical content of the gauged action and the ILST actions are the same.\footnote{See \cite{Gustafsson:1994kr} for a related discussion.} In this paper we establish the latter by a thorough analysis, as well as a more complete analysis of physical configurations of null strings.

\subsection{Equations of motion and constraints}

${\cal V}^a, {\cal W}_a$ and $X^\mu$ are dynamical fields of the gauged action \eqref{gauged_action}. Among these ${\cal V}^a, {\cal W}_a$ are gauge degree of freedom (DoF), as they have vanishing conjugate momentum. As such, the corresponding equations of motion (EoM) are constraints, see \cite{Sheikh-Jabbari:2025tkh} and references therein for a concise discussion. The constraint equations are \cite{Sheikh-Jabbari:2026vqh},
\begin{subequations}
\begin{align}
    \text{EoM for } {\cal V}^a: &\qquad  \frac{\delta S}{\delta \mathcal{V}^a}  = P\cdot D_a X  =0 \qquad (a = 0,1), \label{eq:eom_V}
\\
    \text{EoM for } {\cal W}_a: &\qquad    \frac{\delta S}{\delta {\cal W}_a} =  {\cal V}^a\ {\cal C}_3=0, \qquad {\cal C}_3:= P\cdot X.    \label{eq:eom_W}
\end{align}
\end{subequations}
where $Y\cdot Z=Y^\mu Z_\mu$ and 
\begin{equation}\label{P-mu-def}
    P^\mu:=\kappa \; \mathcal{V}^a  D_a X^\mu\,.
\end{equation}
We crucially note that $P\cdot D_a X= P\cdot \partial_a X+ {\cal C}_3 A_a$. Therefore, the set of constraints we are dealing with takes the form:
\begin{equation}\label{constraints-CI}
 \inbox{  {\cal C}_a=  P\cdot \partial_a X  =0, \qquad 
{\cal C}_3:= P\cdot X=0.    }
\end{equation}
That is, the set of constraints are independent of ${\cal W}_a$ gauge fields. 

Besides the above constraints, we have the EoM for $X^\mu$:
\begin{equation}
D_a\!\left( \mathcal{V}^a \mathcal{V}^b D_b X_\mu \right)= \partial_a \left( \mathcal{V}^a \mathcal{V}^b D_b X_\mu \right) + {\cal W}_a \mathcal{V}^a \mathcal{V}^b D_b X_\mu = 0.
\label{eq:eom_X}
\end{equation}

\subsection{Gauge-fixings and associated constraints}

\paragraph{Fixing the Carroll-Weyl gauge symmetry.}

The gauged action reduces to ILST action upon fixing the $\chi$-gauge symmetry. To this end, we need a diffeomorphism invariant gauge fixing condition that respects the Carrollian structure. The natural choice is,
\begin{equation}
    {\cal V}^a {\cal W}_a=0 \qquad \longrightarrow\quad  P^\mu={\cal V}^a \partial_a X^\mu,
     \ \ D_a {\cal V}^a=\partial_a {\cal V}^a
\end{equation}
This gauge fixing leaves us with residual gauge transformations, defined by the condition \cite{Sheikh-Jabbari:2026cnj, Sheikh-Jabbari:2026vqh},
\begin{equation}\label{chi-Va}
    {\cal V}^a \partial_a \chi=0.
\end{equation}
It is apparent that upon  ${\cal V}^a {\cal W}_a=0$ the gauged action \eqref{gauged_action} reduces to \eqref{ILST}. 
We note that the set of constraints \eqref{constraints-CI} is independent of the specific gauge-fixing adopted here. 

\paragraph{Diffeomorphism gauge-fixing.} The worldsheet diffeomorphism invariance can be used to eliminate or to fix ${\cal V}^a$ DoF. A convenient choice is the \textit{temporal gauge}, in which
\begin{equation}\label{temporal_gauge}
\mathcal{V}^a = (1, 0),\qquad {\cal V}^\tau=1,\ {\cal V}^\sigma=0.
\end{equation}
Recalling \eqref{Va-eta-transf}, $\delta_{\eta} {\cal V}^a=0$ and  \eqref{chi-Va} leaves us with residual $\eta$ transformations $\eta=(\zeta^a, \chi)$, with \cite{Sheikh-Jabbari:2026cnj}
\begin{equation}\label{residual-sym}
    \zeta^a= \left[h(\sigma)+(f'(\sigma)-2\chi(\sigma))\tau\right] \partial_\tau + f(\sigma) \partial_\sigma,  \qquad \chi=\chi(\sigma),
\end{equation}
where $f'=\partial_\sigma f$. The residual diffeos are hence specified by \textit{three} functions of $\sigma$ only and under which, 
\begin{subequations}\label{X-P-under-residual}
\begin{align}
   \delta_\eta X^\mu &= \left[h+(f'-2\chi)\tau\right] \dot{X}^\mu+ f X^\mu{}'+\chi X^\mu, \label{X-under-eta}\\
   \delta_\eta P_\mu &= \left[h+(f'-2\chi)\tau\right] \dot{P}_\mu+ f P_\mu{}'+(f'-\chi) P_\mu, \label{P-under-eta}
\end{align}
\end{subequations}
where $\dot{X}=\partial_\tau X$ and in \eqref{P-under-eta} we have used the fact that $P^\mu={\cal V}^a\partial_a X^\mu$. We note that while under residual transformations \eqref{residual-sym}, $\delta_\eta {\cal W}_a\neq 0$, one can readily verify that ${\cal V}^a \delta_\eta {\cal W}_a=0$ (as a consequence of $\delta_\eta {\cal V}^a=0$, ${\cal V}^a {\cal W}_a=0$). Note also that  $\delta_\eta P_\mu= \zeta^a\partial_a P_\mu + \frac{1}{2}(\partial_a\zeta^a)P_\mu$, i.e it transform as scalar density with weight 1/2, which is expected recalling \eqref{P-mu-def} and that ${\cal V}^a, X^\mu$ have weights 1/2, 0 respectively.

To summarize so far, after ${\cal V}^a, {\cal W}_a$ gauge-fixings these two completely drop out of the dynamics and we remain with residual gauge transformations specified by three functions of $h(\sigma), f(\sigma), \chi(\sigma)$, and three constraints \eqref{constraints-CI}. That is, solutions of EoM, which in the temporal gauge take the form,
\begin{equation}
    P^\mu:=\dot X^\mu, \qquad \dot{P}^\mu=0, 
\end{equation}
are defined up to three freedoms given in \eqref{X-P-under-residual}, and subject to constraints 
\begin{equation}\label{constraints-temporal}
    {\cal C}_1= P^2, \qquad {\cal C}_2= P\cdot X',\qquad {\cal C}_3= P\cdot X\,.
\end{equation}
Upon EoM, the constraints have a simple dynamics:
\begin{equation}
    \dot{\cal C}_1=0,\qquad \dot{\cal C}_2=\frac12 {\cal C}_1',\qquad \dot{\cal C}_3={\cal C}_1\,.
\end{equation}

\subsection{Classical solutions and constraints }

Field equations of null strings are quite simple $\ddot{X}^\mu=0$, with the general solution:
\begin{equation}\label{general_solution}
X^\mu(\tau,\sigma) = \frac{1}{\kappa} {P^\mu(\sigma)}\tau + X_0^\mu(\sigma).
\end{equation}
Under the residual gauge transformations generated by $\eta$,  $\delta_\eta P_\mu$ is given in \eqref{P-under-eta} and 
\begin{equation}
    \delta X_0^\mu= f \ X_0^\mu{} '+\chi X_0^\mu,
\end{equation}
where we used the fact that $\delta \tau =h(\sigma)$. 
The constraints \eqref{constraints-CI} reduce to three $\tau$-independent constraints:
\begin{align}
\texttt{C}_1(\sigma) &:=  {P^\mu(\sigma) P_\mu(\sigma)} = 0, \label{C1} \\
\texttt{C}_2(\sigma) &:=  P_\mu(\sigma) X_0'^\mu(\sigma) =0 , \label{C2}\\
\texttt{C}_3(\sigma) &:= P_\mu(\sigma) X_0^\mu(\sigma)=0 \label{C3}
\end{align}
where $X' = \partial_\sigma X$. While $\texttt{C}_1(\sigma), \texttt{C}_2(\sigma)$ has been considered in the null string literature, e.g. see \cite{Bagchi:2026wcu} and references therein, $\texttt{C}_3(\sigma)$ has been overlooked \cite{Sheikh-Jabbari:2026cnj}. In the next subsection we solve EoM considering the last constraint.

\subsection{Light-cone gauge and residual symmetries}

To obtain physically distinct null string configurations we should impose the three constraints as well as fixing the residual symmetries. Solutions in \eqref{general_solution} involve $2D$ functions of $\sigma$. We note that the $X^\mu, P_\mu$ configurations mapped to each other under residual symmetries \eqref{X-P-under-residual} are gauge, and hence physically, equivalent. So, using the residual symmetries one can  eliminate 3 functions of $\sigma$ by the choice of three functions $h(\sigma), f(\sigma), \chi(\sigma)$. Moreover, physical configurations should satisfy the three $\texttt{C}_1(\sigma), \texttt{C}_2(\sigma), \texttt{C}_3(\sigma)$ constraints, implying that three of $2D$ functions of $\sigma$, $X_0^\mu(\sigma), P_\mu(\sigma)$, can be solved in terms of the remaining ones. Therefore, a generic null string physical configuration  involves $2D-(3+3)=2(D-3)$ functions of $\sigma$. 

To explicitly impose the 3+3 constraints, a convenient way is to fix the light-cone gauge, as is usually performed in the standard tensile string theory \cite{Green:1987sp, Polchinski:1998rq}. To this end, we introduce light-cone coordinates in the target space:
\begin{align}
X^{\pm} = \frac{X^0 \pm X^{D-1}}{\sqrt{2}}, \qquad X^1,\qquad 
X^I \quad (1 = 2, \dots, D-1) \quad \text{(``transverse'')}.
\end{align}
Note that there are  $D-3$ ``transverse coordinates'', in contrast to the standard light-cone coordinates which counts $X^1$ and $X^I$ together as transverse.  The conventional choice  for the {light-cone gauge},
\begin{equation}\label{lightcone_gauge}
X^+ = \frac{p^+}{\kappa} \tau, \qquad P^+ = p^+ \neq 0\quad \text{(constant)}.
\end{equation}
Recalling \eqref{X-P-under-residual}, this fixes $h=0, \chi=f'$. With these choices we can solve for $X^-_0{}^\mu, P^-_\mu$ in terms of the other $X^I_0, P_I$ using $\texttt{C}_1(\sigma)=0, \texttt{C}_3(\sigma)=0$: 
\begin{align}\label{C1-C3-solution-X-}
 P^- = \frac{(P_1)^2}{2p^+}+\frac{(P_I)^2}{2p^+},\qquad  X_0^- = \frac{P_1 X_0^1}{p^+}+ \frac{P_I X_0^I}{p^+} . 
\end{align}

So, we remain with a single residual gauge symmetry function $f(\sigma)$, under which for on-shell configurations, 
\begin{equation}\label{X-P-under-residual--f}
   \delta_f X_0^\mu = (f X_0^\mu{})', \qquad 
   \delta_f P_\mu(\sigma) =  f P_\mu{}', 
\end{equation}
and the constraint $\texttt{C}_2(\sigma)=0$ to be imposed which noting  that $\texttt{C}_2(\sigma)=0$, upon using $\texttt{C}_3(\sigma)=0$, can be recast as $P'\cdot X_0=0$. Using the residual $f$-symmetry, we can fix $X_0^1, P_1$ or a combination thereof. Two particular (while still generic) choices are fixing $P_1(\sigma)$ to a non-constant given function or,  setting $X_0^{1}=x_0^1=$const. (and $x_0^1\neq 0$). 
\begin{itemize}
    \item \textbf{$X_0^{1}=x_0^1=$const. (and $x_0^1\neq 0$)}. This choice leaves us with $f'=0$  sector and
\begin{equation}\label{P1-LCG}
    P'_1=-\frac{1}{x_0^1}  P'_I X_0^I\qquad \Longrightarrow\qquad P_1=-\frac{1}{x_0^1} \int^\sigma P'_I X_0^I+ p_1
\end{equation}

    \item  \textbf{$P^{1}(\sigma)= p_1(\sigma)$ (and $p'_1\neq 0$).} This choice completely fixes $f$ and,
    \begin{equation}\label{X1-LCG}
    X^0_1(\sigma)=-\frac{1}{p'^1}  P'_I X_0^I
\end{equation}
However, since for a closed string $p_1(\sigma)$ is a periodic function, $p'_1(\sigma)$ is a periodic function which averages to zero. It hence always has zeros in $\sigma\in [0,2\pi]$ range. If we denote these zeros by $\sigma_k$, \eqref{X1-LCG} is well-defined if $P'_I X_0^I$ also vanishes at $\sigma_k$.
\end{itemize}

Thus, classical physical configurations are completely specified by $X_0^I, P_I, I=2,\cdots, D-1$. The only remaining part of the residual gauge symmetry for $X_0^1=$const. branch, is constant $f$, $f'=0$, part, corresponding to rigid translations along $\sigma$. For the $P_1(\sigma)=p_1(\sigma)$ branch we do not have a residual $f$-symmetry. For the closed strings that we mainly consider in this paper, this part is absent due to periodic boundary conditions. So, this completes the gauge-fixing and solutions of constraints.

\subsection{Examples of classical null closed string configurations}

For closed strings, the fields are periodic in $\sigma$ with period $2\pi$, up to possible winding. 
The transverse degrees of freedom $P^I(\sigma)$ and $X_0^I(\sigma)$ ($I=2,\dots,D-1$) can therefore be expanded as
\begin{align}
    P^I(\sigma) &= p^I + \sum_{n\neq 0} P_n^I \, e^{in\sigma}, \label{P_expansion} \\
    X_0^I(\sigma) &= x_0^I + w^I \sigma + \sum_{n\neq 0} X_n^I \, e^{in\sigma}, \label{X0_expansion}
\end{align}
where $p^I$, $x_0^I$ are constants (zero modes), $w^I$ are the winding numbers, and the Fourier coefficients satisfy 
$(P_{-n}^I)^* = P_n^I$, $(X_{-n}^I)^* = X_n^I$ for real fields. 
The coordinate $X^1$ is gauge-fixed to a constant $x_0^1$ (no winding) and its momentum $P^1$ is given in terms of $X_0^I, P_I$ as in \eqref{P1-LCG}. As explained in the main text and dealing with tensionless strings, the winding does not contribute to the energy of the string.

The solutions of the previous subsection take the explicit form
\begin{equation}\label{lightcone_gauge-summary}
\begin{split}
X^+ &= \frac{p^+}{\kappa} \tau, \qquad P^+ = p^+ \neq 0\quad \text{(constant)}.\\ 
X^- &= \frac{1}{\kappa^+} P^-(\sigma)\ \tau +X_0^-(\sigma) \\
X^1 &= \frac{1}{\kappa} P^1(\sigma) \ \tau + x_0^1,\qquad P^1(\sigma) = p^1 + \sum_{m\neq 0} P_m^1 \, e^{im\sigma}\\ 
X^I &= \frac{1}{\kappa} P^I(\sigma) \ \tau + X_0^I(\sigma), \qquad I=2,\cdots, D-1
\end{split}
\end{equation}
where 
\begin{equation}
    \begin{split}
     P_m^1 &= \frac{i}{x_0^1}\ \sum_{m\neq 0} \ \frac{n}{m} P^I_n X_{m-n}^I\\
P^-(\sigma)&= \frac1{2p^+}(p_1^2+p_I^2+\sum_{m\neq 0} P^1_{-m} P^1_m+P^I_{-m} P^I_m)+ \sum_{n\neq 0} P_n^- e^{in\sigma},\quad  \\ 
X_0^-(\sigma) &= \frac1{p^+}\big[p_Ix_0^I+p_1 x_0^1+(p_I w^I)\sigma+ \sum_{n\neq 0} X^-_n e^{in\sigma}\big],
    \end{split}
\end{equation}
with
\begin{equation}
\begin{split}
 X^-_n=p_I X_n^I+x_0^I P_I{}_n+\sum_m(1+i\frac{m}{n}) P_m^I X_{n-m}^I, \\   P_n^-=\frac{1}{p^+}P^I_m\big(p_I\delta_{m,n}+\frac{p^1}{x_0^1}\frac{m}{n} X_{n-m}^I\big)+\frac1{2p^+}(M^{IJ}_{mrs})_n P^I_r P^J_s,\\
   (M^{IJ}_{mrs})_n= \delta^{IJ} \delta_{m,r}\delta_{n,r+s}+\frac{rs}{(x_0^1)^2}  X^I_{m-r} X^J_{n-s-m}
\end{split}\end{equation}

We end this part by some comments on the general closed string physical configurations discussed above.
\begin{itemize}
    \item As we see, $X^-$ can have the winding number $w^-=p_I w^I/p^+$.
    \item  Periodicity of $X^-$ (up to winding) implies
\begin{equation}\label{w-P-orth}
    w_I P^I_n=0 \qquad \forall n\neq 0\,.
\end{equation}
\item  Here, we have used $\texttt{C}_3$ to solve for $X_0^-$. This is in contrast to solving for  $X_0^-{}'$ through solving $\texttt{C}_2=0$ that we have for the standard tensile string in the lightcone gauge \cite{Green:1987sp}. The latter means that for the closed strings $\int_0^{2\pi} \text{d}\sigma X^-{}'$ should be $p^+ w^-$; the standard level matching condition. So, in the null string case we do not have level matching condition, while we have $w^-=p_I w^I/p^+$ and \eqref{w-P-orth}.
\end{itemize}

\section{Phase space and canonical formulation}\label{sec:phase_space_total}

To perform a systematic Dirac--Bergmann constraint analysis \cite{Henneaux:1992} of the gauged scale-invariant null string theory, we lift the Lagrangian formulation to the canonical phase space framework. The configuration space of the system at any given worldsheet time $\tau$ is parameterised by the set of local fields:
\begin{equation}
q^A(\tau,\sigma) = \left\{ X^\mu(\tau,\sigma),\; \mathcal{V}^a(\tau,\sigma) ,\; {\mathcal W}_a(\tau,\sigma) \right\},
\label{eq:config_space}
\end{equation}
where $\mu = 0, \dots, D-1$ denotes the target-space spacetime index, and $a \in \{0,1\}$ corresponds to the worldsheet coordinate indices $\{\tau, \sigma\}$. The corresponding canonical momentum densities $p_A(\tau,\sigma)$ conjugate to each configuration field variable are defined via the standard functional variations with respect to the velocity fields:
\begin{equation}
p_A(\tau,\sigma) := \frac{\partial \mathcal{L}}{\partial \dot{q}^A(\tau,\sigma)}.
\label{eq:mom_def}
\end{equation}

\subsection{Canonical conjugate momenta and fundamental Poisson brackets}

The gauged density Lagrangian density $\mathcal{L}$ enters the velocity sector exclusively through the temporal component of the Carroll--Weyl covariant derivative, $D_0 X^\mu = \dot{X}^\mu + {\mathcal W}_0 X^\mu$. Varying the Lagrangian with respect to $\dot{X}^\mu$ isolates the canonical spacetime momentum density:
\begin{equation}
P_\mu(\tau,\sigma) := \frac{\partial \mathcal{L}}{\partial \dot{X}^\mu} = \kappa \, \mathcal{V}^0 \mathcal{V}^a D_a X_\mu = \kappa \mathcal{V}^0 \left[ \mathcal{V}^0 (\dot{X}_\mu + \mathcal{W}_0 X_\mu) + \mathcal{V}^1 D_1 X_\mu \right]
\label{eq:target_momentum}
\end{equation}
This relation defines the physical momentum configuration of the string embedding sector before the imposition of any spatial gauge choices or on-shell reductions. To have a non-vanishing momentum, we assume that $\mathcal{V}^0\neq 0$.  

The generalised action~\eqref{gauged_action} does not contain explicit worldsheet time derivatives of $\mathcal{V}^a$ and the Carroll--Weyl gauge connection ${\mathcal W}_a$ and the velocity fields $\dot{\mathcal{V}}^a$ and $\dot{\mathcal{W}}_a$ are algebraically absent from the Lagrangian density. Consequently, the canonical momentum densities conjugate to these auxiliary fields vanish identically across the entire phase space:
\begin{equation}
\Pi_a(\tau,\sigma) := \frac{\partial \mathcal{L}}{\partial \dot{\mathcal{V}}^a} = 0, \qquad
{\mathcal P}^a(\tau,\sigma) :=\frac{\partial \mathcal{L}}{\partial \dot{\mathcal{W}}_a} = 0 .
\label{eq:Pi_W_original}
\end{equation}
Thus, as in any gauge theory, we are dealing with a constrained system \cite{Henneaux:1992} and $\mathcal{V}^a$ and  ${\mathcal W}_a$  act as Lagrange multipliers rather than true physical degrees of freedom.

\paragraph{Fundamental Poisson brackets.} The symplectic structure on the unreduced phase space is established by introducing the standard equal-time canonical Poisson brackets. The non-vanishing equal-time relations among the configuration fields and their conjugate momentum densities are defined as follows:
\begin{align}
\left\{X^\mu(\tau,\sigma), \, P_\nu(\tau,\sigma')\right\} &= \delta^\mu_\nu \; \delta(\sigma - \sigma'), \label{eq:pb_x_p} \\
\left\{\mathcal{V}^a(\tau,\sigma), \, \Pi_b(\tau,\sigma')\right\} &= \delta^a_b \; \delta(\sigma - \sigma'), \label{eq:pb_v_pi} \\
\left\{{\mathcal W}_a(\tau,\sigma), \, {\mathcal P}^b(\tau,\sigma')\right\} &= \delta_a^b \; \delta(\sigma - \sigma'). \label{eq:pb_w_p}
\end{align}
All other cross-combinations of phase space field variables vanish identically at equal worldsheet times.

\subsection{Canonical framework and Legendre transformation}

The identical vanishing of the auxiliary momentum densities~\eqref{eq:Pi_W_original} implies that the phase space trajectories are restricted to a subspace defined by a set of primary constraints in the sense of Dirac. We denote these constraints as:
\begin{subequations}
\begin{align}
 \Pi_a(\tau,\sigma) &\approx 0, \label{eq:primary_phi} \\
 {\mathcal P}^a(\tau,\sigma) &\approx 0, \label{eq:primary_psi}
\end{align}
\end{subequations}
where the symbol $\approx$ denotes weak equality on the primary constraint surface. $\Pi_a(\tau,\sigma), {\mathcal P}^a(\tau,\sigma)$ are four independent primary constraints per worldsheet point which must be systematically preserved under the time evolution generated by the canonical Hamiltonian.

With the symplectic structure and primary constraints established, we construct the canonical Hamiltonian $H_c$ via the standard Legendre transformation from the Lagrangian density $\mathcal{L}$ \cite{Henneaux:1992}:
\begin{equation}
H_c = \int d\sigma \; \left( P_\mu \dot{X}^\mu + \Pi_a \dot{\mathcal{V}}^a + \mathcal{P}^a \dot{\mathcal{W}}_a - \mathcal{L} \right),
\label{eq:H_c_Legendre}
\end{equation}
where the integration is performed over the spatial coordinate $\sigma$ of the  worldsheet. To map this expression explicitly onto the phase space variables, we should replace $\dot{X}^\mu$ with the target-space momentum density $P_\mu$ \eqref{eq:target_momentum}, yielding 
\begin{equation}
\dot{X}_\mu = \frac{P_\mu}{\kappa (\mathcal{V}^0)^2} - \mathcal{W}_0 X_\mu - \frac{\mathcal{V}^1}{\mathcal{V}^0} D_1 X_\mu .
\label{eq:X_dot_inverted}
\end{equation}
Substituting the explicit velocity profile~\eqref{eq:X_dot_inverted} back into the Legendre definition~\eqref{eq:H_c_Legendre}, the canonical Lagrangian density $\mathcal{L} = \frac{\kappa}{2} \mathcal{V}^a \mathcal{V}^b D_a X \cdot D_b X$ is recast entirely in terms of phase space variables. After a straightforward algebra the canonical Hamiltonian density is obtained as,
\begin{equation}
\inbox{H_c = \int d\sigma \; \left[ \frac{P_\mu P^\mu}{2\kappa (\mathcal{V}^0)^2} - \frac{\mathcal{V}^1}{\mathcal{V}^0} P_\mu D_1 X^\mu - \mathcal{W}_0 (P_\mu X^\mu) \right]} 
\label{eq:H_c_final}
\end{equation}
\section{Total Hamiltonian and stability of the constraint manifold}\label{sec:total_stability}

To generate the full classical dynamics on the constrained phase space, we define the total Hamiltonian $H_T$ by appending the complete set of primary constraints $\{\Pi_a, {\cal P}^a\}$ to the canonical functional $H_c$ using  Lagrange multipliers $\lambda^a(\sigma)$ and $\mu_a(\sigma)$ \cite{Henneaux:1992}
\begin{equation}
H_T = H_c + \int d\sigma \; \left[ \lambda^a(\sigma)\ \Pi_a(\sigma) + \mu_a(\sigma)\ {\cal P}^a(\sigma) \right].
\label{eq:H_T_definition}
\end{equation}
Consistency of the Dirac-Bergmann algorithm requires that all primary constraints be rigorously preserved under the temporal evolution governed by $H_T$. For any local primary constraint, this consistency condition takes the form of a stability equation $\dot{\Phi} = \{\Phi, H_T\} \approx 0$, which must hold weakly on the constraint surface. $\{\Pi_a, \mathcal{P}^a\}$, they trivially commute among themselves under the canonical Poisson brackets~\eqref{eq:pb_v_pi} and~\eqref{eq:pb_w_p}. Consequently, the temporal stability of the primary sector reduces to the weak vanishing of their brackets with the canonical Hamiltonian $H_c$:
\begin{equation}
\dot{\Pi}_a = \{\Pi_a, H_c\} \approx 0, \qquad \dot{{\cal P}}^a = \{{\cal P}^a, H_c\} \approx 0.
\label{eq:primary_stability_conditions}
\end{equation}
As the next steps we should study the secondary constraints and closure of the chain of (secondary) constraints.

\paragraph{Secondary constraints associated with the primary constraints $\Pi_a(\sigma) \approx 0$.} Computing the canonical Poisson brackets using the functional identity $\{\Pi_a(\sigma), H_c\} = -\delta H_c / \delta \mathcal{V}^a(\sigma)$, we find,
\begin{equation} \label{eq:phia_bracket}
\dot{\Pi}_0(\sigma) = \frac{1}{(\mathcal{V}^0)^2} \left[ \frac{P^2}{\kappa \mathcal{V}^0} - \mathcal{V}^1 P_\mu D_1 X^\mu \right] \approx 0, \qquad 
\dot{\Pi}_1(\sigma) = \frac{1}{\mathcal{V}^0} P_\mu D_1 X^\mu \approx 0. 
\end{equation}
The above yields, 
\begin{subequations}
\begin{align}
{\mathcal C}_1 := P_\mu P^\mu \approx 0.
\label{eq:C1_def}\\
{\mathcal C}_2 := P_\mu D_1 X^\mu \approx 0.\label{eq:C2_def} 
\end{align}
\end{subequations}
The emergence of $\mathcal{C}_1$ and $\mathcal{C}_2$ matches the standard diffeomorphism and reparameterization structures expected in Carrollian string models \cite{Sheikh-Jabbari:2026cnj}.


\paragraph{Secondary constraint associated with $\mathcal{P}^a(\sigma) \approx 0$.} Computing the canonical variations via the functional relations $\{{\cal P}^a(\sigma), H_c\} = -\delta H_c / \delta \mathcal{W}_a(\sigma)$, yields
\begin{align}\label{eq:psia_bracket}
\dot{{\cal P}}^0(\sigma) = P_\mu(\sigma) X^\mu(\sigma) \approx 0,  \qquad
\dot{{\cal P}}^1(\sigma) = \frac{\mathcal{V}^1(\sigma)}{\mathcal{V}^0(\sigma)} P_\mu(\sigma) X^\mu(\sigma) \approx 0. 
\end{align}
Both relations yield a single secondary constraint profile across the worldsheet, which manifests explicitly as the internal scaling (dilatation) generator \cite{Sheikh-Jabbari:2026cnj},
\begin{equation}
{\mathcal C}_3 := P_\mu X^\mu \approx 0.
\label{eq:C3_def}
\end{equation}
The constraint $\mathcal{C}_3$ encodes the local scaling symmetry of the embedding sector and is a structural cornerstone of the full classical theory.

\paragraph{Closure of the constraint chain.} To establish the complete consistency of the Dirac-Bergmann chain, we should verify the temporal stability of the newly discovered secondary constraints $\{{\mathcal C}_1, {\mathcal C}_2, {\mathcal C}_3\}$ under the evolution generated by the total Hamiltonian,
\begin{equation}
\dot{\mathcal{C}}_i(\sigma) = \{{\mathcal C}_i(\sigma), H_T\} \approx 0 \quad \text{for} \quad i \in \{1,2,3\}.
\label{eq:secondary_stability}
\end{equation}
The secondary constraints are functions solely of the coordinates $X^\mu$ and their conjugate momenta $P_\mu$, they commute identically with all primary constraints, which depend exclusively on the auxiliary momenta. Thus, the evaluation of \eqref{eq:secondary_stability} reduces to computing their canonical Poisson brackets with the canonical Hamiltonian $H_c$. Direct functional computation yields:
\begin{align}
\dot{\mathcal C}_1 &= \{{\mathcal C}_1, H_c\} \approx 0, \label{eq:C1_stable} \\
\dot{\mathcal C}_2 &= \{{\mathcal C}_2, H_c\} =\frac12 D_1 {\cal C}_1 \approx 0, \label{eq:C2_stable} \\
\dot{\mathcal C}_3 &= \{{\mathcal C}_3, H_c\} =\frac12 \dot{\cal C}_1\approx 0. \label{eq:C3_stable}
\end{align}
That is, all stability brackets vanish weakly on the secondary constraint surface without requiring the introduction of further algebraic conditions. So, the chain of secondary constraints terminate with the inclusion of ${\cal C}$'s. The system achieves full classical closure: the complete set of seven constraints per worldsheet point, given by $\{\Pi_0, \Pi_1, {\cal P}^0, {\cal P}^1, \mathcal{C}_1, \mathcal{C}_2, \mathcal{C}_3\}$, forms a stable, closed constraint manifold that completely governs the canonical dynamics of the gauged null string.

\section{Constraint algebra}\label{sec:constraint_algebra}

To establish the ultimate physical properties and structural consistency of the gauged scale-invariant null string, we evaluate the canonical Poisson bracket algebra among the complete set of primary and secondary constraints. This step determines whether the constraint manifold generates a closed gauge algebra and dictates the systematic sorting of the system into first-class or second-class sectors under the Dirac classification scheme.

\subsection{ Algebra of smeared secondary constraints}

To manipulate the underlying algebraic structures associated with Poisson bracket of the constraints rigorously, we map the constraints $\{{\mathcal C}_1, {\mathcal C}_2, {\mathcal C}_3\}$ to globally well-defined functionals by smearing them against smooth, independent, periodic test functions $f(\sigma), g(\sigma), k(\sigma) $:
\begin{equation}
\boldsymbol{\mathcal{C}}_1[f] := \int d\sigma \, f(\sigma) {\mathcal C}_1(\sigma), \quad
\boldsymbol{\mathcal{C}}_2[g] := \int d\sigma \, g(\sigma) {\mathcal C}_2(\sigma), \quad
\boldsymbol{\mathcal{C}}_3[k] := \int d\sigma \, k(\sigma) {\mathcal C}_3(\sigma).
\label{eq:smeared_definitions}
\end{equation}
Evaluating the equal-time canonical brackets among these integrated functionals using the fundamental relations~\eqref{eq:pb_x_p}, a functional computation establishes the classical closed constraint algebra:
\begin{subequations}\label{const-algebra-smeared}
\begin{align}
\left\{\boldsymbol{\mathcal{C}}_1[f], \, \boldsymbol{\mathcal{C}}_1[g]\right\} &= 0, \label{eq:algebra_c1_c1} \\
\left\{\boldsymbol{\mathcal{C}}_1[f], \, \boldsymbol{\mathcal{C}}_2[g]\right\} &= \,\boldsymbol{\mathcal{C}}_1[ fg'-f' g], \label{eq:algebra_c1_c2} \\
\left\{\boldsymbol{\mathcal{C}}_1[f], \, \boldsymbol{\mathcal{C}}_3[g]\right\} &= -2\,\boldsymbol{\mathcal{C}}_1[ f g], \label{eq:algebra_c1_c3} \\
\left\{\boldsymbol{\mathcal{C}}_2[f], \, \boldsymbol{\mathcal{C}}_2[g]\right\} &= \boldsymbol{\mathcal{C}}_2[ f g' - f' g], \label{eq:algebra_c2_c2} \\
\left\{\boldsymbol{\mathcal{C}}_2[f], \, \boldsymbol{\mathcal{C}}_3[g]\right\} &= \boldsymbol{\mathcal{C}}_3[ f g'], \label{eq:algebra_c2_c3} \\
\left\{\boldsymbol{\mathcal{C}}_3[f], \, \boldsymbol{\mathcal{C}}_3[g]\right\} &= 0, \label{eq:algebra_c3_c3}
\end{align}
\end{subequations}
where prime denotes the spatial derivative with respect to $\sigma$ (e.g., $g' = \partial_\sigma g$). 

The resulting Lie-Poisson structure yields crucial geometric insights into the underlying symmetries of the theory:
\begin{itemize}
\item Eq.~\eqref{eq:algebra_c2_c2} shows that the spatial diffeomorphism generator $\boldsymbol{\mathcal{C}}_2[f]$ forms the standard Witt or 1D vector field algebra ($\text{Diff}(S^1)$) on the spatial circle of the worldsheet.
\item $\boldsymbol{{\cal C}}_1, \boldsymbol{{\cal C}}_2$ define a BMS$_3$ algebra. We crucially note that $\boldsymbol{{\cal C}}_3$ is not an ideal of the BMS$_3$ cf. \eqref{eq:algebra_c2_c3}. Therefore, overlooking $\boldsymbol{{\cal C}}_3$, as was done in the null string literature prior to \cite{Sheikh-Jabbari:2026cnj},  yields an inconsistent and incomplete treatment of the null string theory.
\item Eqs.~\eqref{eq:algebra_c1_c2} and~\eqref{eq:algebra_c1_c3} establish that the mass-shell constraint $\boldsymbol{\mathcal{C}}_1$ transforms covariantly as a conformal primary field of weight $2$ under spatial reparameterizations, while $\boldsymbol{\mathcal{C}}_3$ is a primary of weight $1$ under the Witt algebra. 

\item Eq.~\eqref{eq:algebra_c1_c3} outlines the non-trivial coupling between mass-shell invariance and local Carroll-Weyl scalings. In particular, the factor $-2$ shows that $\boldsymbol{\mathcal{C}}_1$ transforms with weight $2$ under the Carroll-Weyl scaling generated by $\boldsymbol{\mathcal{C}}_3$. Under the same scaling, $\boldsymbol{\mathcal{C}}_2$ is not a scaling primary, as seen from the transformations, 
\begin{equation}\label{chi-C2-C3}
    \delta_\chi P_\mu(\sigma) =-\chi P_\mu, \quad  \delta_\chi X^\mu(\sigma)=+\chi X^\mu \quad \Longrightarrow \quad \delta_\chi {\cal C}_2=\chi' {\cal C}_3,\quad \delta_\chi {\cal C}_3=0, \quad \delta_\chi {\cal C}_1=-2\chi {\cal C}_1.
\end{equation}

\item  The algebra \eqref{const-algebra-smeared} is invariant under $\boldsymbol{\mathcal{C}}_1\to \lambda \boldsymbol{\mathcal{C}}_1$, for any arbitrary number $\lambda$. This is a key feature of Carrollian structure and the so-called supertranslations (generated by $\boldsymbol{\mathcal{C}}_1$).

\end{itemize}

\subsection{Constraint algebra is Extended \texorpdfstring{BMS$_3$}{BMS3}}

As mentioned, \eqref{const-algebra-smeared} is BMS$_3$ algebra extended by $\boldsymbol{{\cal C}}_3$ generators. To see this in a more familiar form, we consider Fourier modes of the constraints, we take $f, g, k$ to be Fourier basis functions $e^{-in\sigma}$, explicitly, 
\begin{equation}
L_n =  \int_0^{2\pi} d\sigma \, e^{-in\sigma} \mathcal{C}_2(\sigma), \quad
M_n = \int_0^{2\pi} d\sigma \, e^{-in\sigma} \mathcal{C}_1(\sigma), \quad
S_n = \int_0^{2\pi} d\sigma \, e^{-in\sigma} \mathcal{C}_3(\sigma).
\label{eq:mode_integrals}
\end{equation}
The canonical brackets evaluate to a closed, infinite-dimensional classical mode algebra:
\begin{subequations}
\begin{align}
\{L_m, L_n\} &= i (m-n) L_{m+n}, \label{eq:mode_L_L} \\
\{L_m, M_n\} &= i (m - n) M_{m+n}, \label{eq:mode_L_M} \\
\{M_m, M_n\} &= 0, \label{eq:mode_M_M} \\
\{L_m, S_n\} &= i n S_{m+n}, \label{eq:mode_L_N} \\
\{M_m, S_n\} &= -2 M_{m+n}, \label{eq:mode_M_N} \\
\{S_m, S_n\} &= 0, \label{eq:mode_S_n}
\end{align}
\end{subequations}
with the standard mode reality conditions $L_{-n} = L_n^*$, $M_{-n} = M_n^*$, and $S_{-n} = S_n^*$.

\paragraph{Comments on the algebra.} 
The subalgebra generated by $L_n$ and $M_n$ is precisely the centerless BMS$_3$ algebra, as seen from \eqref{eq:mode_L_L}, \eqref{eq:mode_L_M}, and \eqref{eq:mode_M_M}. The full algebra extends BMS$_3$ by the generators $S_n$, with the additional brackets \eqref{eq:mode_L_N} and \eqref{eq:mode_M_N}. 

The $M_n, S_m$ generators form  Abelian ideals of the Witt algebra generated by $L_n$, since $\{L_m, M_n\} \propto M_{m+n}$ and $\{L_m, S_n\} \propto S_{m+n}$. The numeric factors, respectively $m-n$ and $n$, show that $M_n$ (supertranslations in the standard BMS$_3$ terminology) has weights $2$ and $S_n$ has weight $1$ under the Witt algebra. The algebra is invariant under $M_n\to \lambda M_n$ scalings and the factor $-2$ in \eqref{eq:mode_M_N} shows that supertranslations have weight $2$ under Carroll-Weyl scalings generated by $\chi$; cf. \eqref{chi-C2-C3}. 

This algebra has appeared in the literature on Carrollian field theories \cite{Batlle:2024ahz} and as the boundary symmetry algebra of three-dimensional Einstein gravity on spacetimes with a null boundary \cite{Adami:2020ugu, Adami:2021nnf}. It admits three central extensions in $LL$, $MM$, $SS$ commutators. Most notably, it does not admit the BMS central extension in the $LM$ commutator \cite{Batlle:2024ahz}. More detailed analysis will be discussed elsewhere.\footnote{We thank Daniel Grumiller for discussions on this point.}

\section{Concluding remarks}

We have established through a thorough Hamiltonian analysis that the three constraints ${\cal C}_1, {\cal C}_2, {\cal C}_3$ are necessary for a consistent treatment of the null string system, as advocated in \cite{Sheikh-Jabbari:2026cnj}. As shown in the  companion paper \cite{Sheikh-Jabbari:2026vqh}, one can fix the Carroll-Weyl gauge symmetry and gauge-fix ${\cal W}_a$ gauge-field; the theory \eqref{gauged_action} is hence gauge equivalent to the ILST action \eqref{ILST}. We have now set a consistent classical stage for the null strings, according which, on a $D$ dimensional Minkowski target space there are $D-3$ propagating string modes. The $\chi$-symmetry and the associated constraint $\mathcal{C}_3 \approx 0$, consistently remove one extra mode compared to the standard tensile string and/or existing null string literature (which have $D-2$ propagating modes). 

The next task is to quantize the system. One can perform canonical  and/or path integral quantizations. For the canonical case, as discussed in \cite{Sheikh-Jabbari:2025tkh, Bagchi:2024tyq, Dutta:2024gkc} one can impose the constraints as ``right-action'' or as ``sandwich conditions''. One should then work through the quantization considering all three constraints in either of right-action or sandwich quantization for the null string. In the path integral, one should introduce a new ghost due to Carroll-Weyl scaling and the textbook $bc$ ghost system \cite{Polchinski:1998rq} should be promoted to ``$bcs$ ghost system''. This consistent quantization scheme may have bearings on the critical dimensions for the null strings, as well as other criteria required by the cancellation of anomalies of all gauge symmetries. Moreover, realization of the Carroll-Weyl scaling will have bearings on the interacting null string and null string vertex operators, that should be studied on their own turn.

\begin{acknowledgments}
We thank Arjun Bagchi, Aritra Banerjee, Daniel Grumiller and Ida Rasulian for discussions and Giulio Bonelli and Bo Sundborg for helpful email exchanges and discussions. MMShJ acknowledges Iranian National Science Foundation (INSF) research chair grant No.40451653. The work of HY is supported in part by Beijing Natural Science Foundation under Grant No. IS23013.
\end{acknowledgments}

\end{document}